\documentclass[twocolumn,english,pre,showpacs]{revtex4}
\usepackage[T1]{fontenc}
\usepackage[latin9]{inputenc}
\usepackage{amsmath}
\usepackage{amssymb}
\usepackage{graphicx}
\usepackage{esint}
\usepackage{amsfonts}
\usepackage{array}
\usepackage{multirow}
\usepackage{amstext}
\usepackage{CJK}
\usepackage{babel}
\usepackage{natbib}
\usepackage{subfigure}

\begin{document}

\nocite{*}

\title{Improved Fluid Perturbation Theory: Equation of state for Fluid Xenon}

\author{Qiong Li$^{1}$\footnotemark[1], Hai-Feng Liu$^{1}$, Gong-Mu Zhang$^{1}$, Yan-Hong Zhao$^{1}$, \\ Ming-Feng Tian$^{1}$, Hai-Feng Song$^{1}$}

\affiliation{$^{1}$Laboratory of Computational Physics, \\Institute
of Applied Physics and Computational Mathematics, Beijing 100094, China\\
\footnotemark[1]Email: liqiong@itp.ac.cn}

\begin{abstract}

The traditional fluid perturbation theory is improved by taking electronic excitations and ionizations into account, in the framework of average ion spheres.
It is applied to calculate the equation of state for fluid Xenon, which turns out in good agreement with the available shock data.

\end{abstract}


\maketitle

\section{Introduction}

The interest in the equation of state (EOS) is stimulated by the applications in, e.g., internal confinement fusion research and in astrophysics \cite{R1,R2}.
Noble gases are intensively studied due to their simple electronic structure as closed shell systems \cite{R3,S1,S2,S3,S4,S5,Root2010,Chen2009,Schwarz2005,Chen2007,Chen2015}.
For the supercritical fluid with the electronic ground state, the thermodynamical property can be well described by the fluid perturbation theory \cite{Redmer2010, BH1976, Ross1979, Kerley1980}. As the temperature or the density rises, electronic excitations and ionizations will take place. For low Z elements such as hydrogen and helium, this region can be well described by the chemical model, regarding the system as an interacting and reacting mixture of molecules, atoms, ionic species and free electrons \cite{Schwarz2005,Chen2007}. Since there is no exact formula for the Coulomb energy of the partially ionized mixture, the Pad\'e interpolation formula for the fully ionized mixture are usually applied, which essentially employs the approximation of ions with bound electrons by point-like particles with the same net charge \cite{Helium}. However, for high Z element Xenon, the sizes of Xe$^{+}$ and Xe$^{++}$ are so large that the point-like approximation becomes seriously unreasonable and thus may bring errors of very big uncertainty.

In this paper, we have carried out a study of the EOS of supercritical fluid Xenon, based on the hard sphere variational formulation of the fluid perturbation theory (FVT). It is found that the 300K pressure curve is in good agreement with the experimental data up to 9.5 g/cc, but the principal Hugoniot pressure curve diverges from the shock data very quickly as the density increases up to 5.6 g/cc. The FVT method is improved by taking electronic excitations and ionizations into account, in the framework of average ion spheres, and is shown to work well in comparison with the shock compression data \cite{Nellis,Urlin,Radousky}.

The rest of the paper is organized as follows. In Sec. II, the models are described. In Sec. III, the results are shown. Finally, summary and outlook is given.

\section{Model Setup}

This section describes the models and proceeds in two steps. First, the FVT model is presented. Then, the FVT model is improved by taking electronic excitations and ionizations into account, using the theory of ionization equilibrium in the framework of average ion spheres.

\subsection{FVT}

In the FVT model, the thermodynamical system is defined by the Helmholtz
free energy function written as

\begin{equation}
F(N,V,T)=F_{\text{id}}+F_{\text{conf}},\label{eq:FVT}
\end{equation}
where $F_{\text{id}}$, $F_{\text{conf}}$ represent
contributions due to the translational kinetic energy and the intermolecular potential energy, respectively.

The kinetic term is given by the Maxwell-Boltzmann statistics for
the molecules,

\begin{equation}
F_{\text{id}}=Nk_{B}T\left[\ln\left(n\Lambda^{3}\right)-1\right],
\end{equation}
with the Boltzmann constant $k_{B}$, the temperature $T$, the number
density $n$, and the thermal de Broglie wavelength $\Lambda=\sqrt{2\pi\hbar^{2}/mk_{B}T}$.

The configurational term $F_{\text{conf}}$ is obtained by the fluid
perturbation theory which treats the strongly repulsive part of the
inter-atomic potential within the hard sphere model and the weak attractive
part as a perturbation. The interactions among Xenon atoms are described
by an effective pair potential of the exp-6 form, which has been successfully
applied to describe the EOS of many materials over a wide range of
densities and temperatures. With the short repulsion
approximated by a hard sphere, we can perturbatively expand the free
energy of the system around the hard sphere system to the first order,
$F_{\text{HS}}+\left\langle \Phi-\Phi_{\text{HS}}\right\rangle _{\text{HS}}$,
where $F_{\text{HS}}$ is the Carnahan-Starling hard sphere free energy
function \cite{CS1969},

\begin{equation}
F_{\text{HS}}=\frac{4\eta-3\eta^{2}}{\left(1-\eta\right)^{2}}Nk_{B}T,
\end{equation}
and the perturbative part is written as

\begin{equation}
\left\langle \Phi-\Phi_{\text{HS}}\right\rangle _{\text{HS}}=\frac{2\pi N^{2}}{V}\int_{d_{0}}^{\infty}g_{\text{HS}}(r,\eta)\Phi(r)r^{2}dr,
\end{equation}
with the pair distribution function of hard spheres $g_{\text{HS}}(r,\eta)$
given by the analytical expression derived from the PercusšCYevick
approximation \cite{Wertheim1963} and the molecular potential $\Phi(r)$
given by
\begin{equation}
\Phi(r)=\begin{cases}
\frac{\varepsilon}{\alpha-6}\left[6\exp\left(\alpha(1-\frac{r}{r_{a}})\right)-\alpha\left(\frac{r_{a}}{r}\right)^{6}\right] & ,\, r\geq W\\
A\exp(-Br) & ,\, r\leq W
\end{cases}
\end{equation}
Following the GibbsšCBogoliubov inequality \cite{MC1970},
\begin{equation}
F_{\text{conf}}\leq F_{\text{HS}}+\left\langle \Phi-\Phi_{\text{HS}}\right\rangle _{\text{HS}}
\end{equation}
the free energy $F_{\text{conf}}$ is determined by minimizing the
right-hand-side with respect to the packing fraction $\eta=\pi d_{0}^{3}N/6V$
with $d_{0}$ the hard sphere diameter.

The fluid perturbation theory developed by the reference system of the inverse 12th-power potential \cite{Ross1979} is formally identical to the original hard sphere
variational formulation described above, except for the additional term given by
\begin{eqnarray}
F_{12}(\eta) & = & -\left(\eta^{4}/2+\eta^{2}+\eta/2\right).
\end{eqnarray}
Consequently, the free energy $F_{\text{conf}}$ is determined by minimizing the inequality
\begin{equation}
F_{\text{conf}}\leq F_{\text{HS}}+F_{12}(\eta)+\left\langle \Phi-\Phi_{\text{HS}}\right\rangle _{\text{HS}}
\end{equation}
with respect to the packing fraction.

The thermodynamical quantities such as pressure, internal energy, and
entropy can be obtained by taking the appropriate derivatives of the
total free energy. With the help of the Hugoniot relation

\begin{equation}
E_{H}=E_{0}+\frac{1}{2}(P_{H}+P_{0})(V_{0}-V_{H})
\end{equation}
the EOS defines the Hugoniot state, which provides a way to check the
accuracy of EOS when compared with the experimental Hugoniot state.

\begin{table}[tbp]
\centering
\begin{tabular}{lcccccc}
\hline
    & $\varepsilon/k_{B}$(K) & $\alpha$ & $r_{a}$(\AA) & $W$(\AA) & $A/k_{B}$(eV) & $B$(\AA$^{-1}$)  \\  \hline
SW  & 235.0 & 13.0 & 4.47 & $-$ & $-$ & $-$  \\  \hline
Fried-228  & 243.1 & 13.0 & 4.37 & $1.284$ & $1625.5$ & $2.071$  \\  \hline
FB  & 246.0 & 12.5 & 4.43 & $1.370$ & $633.8$ & $1.680$  \\  \hline
\end{tabular}
\caption{\label{table:par} Parameters for the Xenon pair potential of the exp-6 form. The "SW" potential is determined by the liquid argon shock wave data using the corresponding states theory \cite{Ross1980} and has been used in Ref.\cite{1658}. The "Fried-228" potential has been used in Refs.\cite{Chen2009, Schwarz2005}. The "FB" potential is fitted by the high-energy atomic-beam scattering data \cite{1658}. }
\end{table}

\begin{figure}
\includegraphics[bb=50 28 700 500, width=8cm]{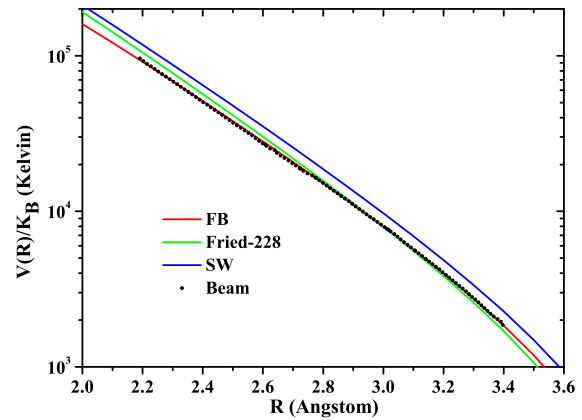}
\caption{\label{fig:potential} (Color online) The Xenon pair potentials as a function of the interatomic separation. "Beam" is derived from the high-energy atomic-beam scattering data \cite{1658}. "SW", "Fried-228" and "FB" are plotted by the exp-6 functions with the parameters given in Table \ref{table:par}. }
\end{figure}

\subsection{IFVT}
For the system of a finite temperature and a finite density, its Helmholtz free energy can be decomposed into three parts - the cold part, the ionic thermal part and the electronic thermal part. In the FVT model, the electronic thermal part is not considered, which includes electronic excitations and ionizations. In the average ion sphere approximation, the fluid perturbation theory is applied to describe the interaction between neutral ion spheres, and the electronic excitations and ionizations can be described using the theory of ionization equilibrium in the average ion spheres \cite{IEQ}. The FVT model improved in the framework of average ion spheres will be referred to as IFVT hereafter.

In the IFVT model, the total Helmholtz free energy can be divided
into two parts,

\begin{equation}
F=(F_{\text{id}}+F_{\text{conf }})+F_{\text{ex}},
\end{equation}
in which the first part $(F_{\text{id}}+F_{\text{conf }})$ is the
same as that of the FVT model, with $\Phi(r)$ interpreted as the
effective pair potential between neutral ion spheres, and the second
part $F_{\text{ex}}$ is contributed by electronic excitations and
ionizations in the average ion sphere.

The total energy inside the ion sphere with $z$ free electrons can
be decomposed into three parts
\begin{eqnarray}
\epsilon_{\text{tot}}(z) & = & \epsilon_{\text{bound}}(z)+\epsilon_{\text{free}}(z)+\epsilon_{\text{b-f}}(z).
\end{eqnarray}
Here, $\epsilon_{\text{bound}}$ is the energy of bound electrons
and given by
\begin{equation}
\epsilon_{\text{bound}}(z)=\epsilon_{z,0}+\varepsilon_{z,\alpha},
\end{equation}
where $\epsilon_{z,0}$ is the ground state energy for an isolated
free ion of charge $z$, and $\varepsilon_{z,\alpha}$ is the excited
state energy relative to the ground state. $\epsilon_{\text{free}}$
is the kinetic energy of free electrons, and $\epsilon_{\text{b-f}}$
is the interaction energy between the bound and free electrons. In
the mean field theory, $\epsilon_{\text{b-f}}$ can be treated as
a shift in the ion energy (continuum lowering) so that it need not
be considered explicitly.

The partition function for an average ion sphere can be obtained by
summing over all configurations of the electrons inside the ion sphere.
First we sum over all the free electron states for a particular configuration
of the bound electrons. In the ion sphere of the $z$th ionization
stage, the kinetic free energy of free electrons obeying Fermi-Dirac
statistics is given by
\begin{equation}
F_{0}(z)=zk_{B}T\left(\xi-\frac{2}{3}\frac{I_{3/2}(\xi)}{I_{1/2}(\xi)}\right),
\end{equation}
where the Fermi integrals are defined by
\begin{equation}
I_{\nu}(\xi)\equiv\int_{0}^{\infty}dx\frac{x^{\nu}}{e^{x-\xi}-1},\qquad(\nu=1/2,\,3/2)
\end{equation}
and the electronic chemical potential $\xi\equiv\mu_{e}^{\text{id}}/k_{B}T$
is determined by
\begin{equation}
I_{1/2}(\xi)=\frac{\sqrt{\pi}}{4}\frac{zN}{V}\lambda_{e}^{3},
\end{equation}
with the electronic wavelength $\lambda_{e}=\frac{h}{\sqrt{2\pi m_{e}k_{B}T}}$.
Next we sum over all stages of ionization and all bound state configurations.
Finally, the partition function of the average ion sphere is obtained as

\begin{eqnarray}
Z_{ex} & = & \sum_{z}\sum_{\alpha}g_{z,\alpha}e^{-(\epsilon_{z,0}+\varepsilon_{z,\alpha}+F_{0}(z))/k_{B}T},
\end{eqnarray}
where $g_{z,\alpha}$ and $\epsilon_{z,0}+\varepsilon_{z,\alpha}$
are the statistical weights and the energy levels of the ions. The second
part of the total free energy is then given by
\begin{equation}
F_{ex}=-Nk_{B}T\ln Z_{ex}.
\end{equation}

\begin{figure}[pbt]
\includegraphics[bb=40 10 700 500, width=8cm]{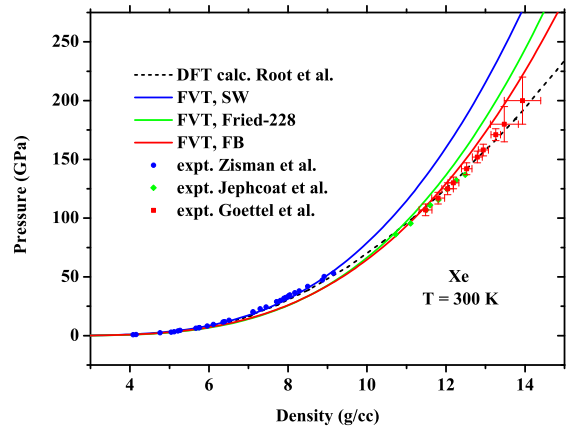}
\caption{\label{fig:298K} (Color online) The pressure-density
relation of xenon at the room temperature. The results calculated by
the FVT model using the pair potentials of "SW", "Fried-228" and
"FB" respectively are compared. The DFT (density-functional theory)
calculation \cite{Root2010} is also shown, together with the
experimental results \cite{Zisman,Jephcoat,Goettel}.  }
\end{figure}

\begin{figure*}[btp]
\begin{center}
\includegraphics[bb = 53 11 722 538,  width=8cm]{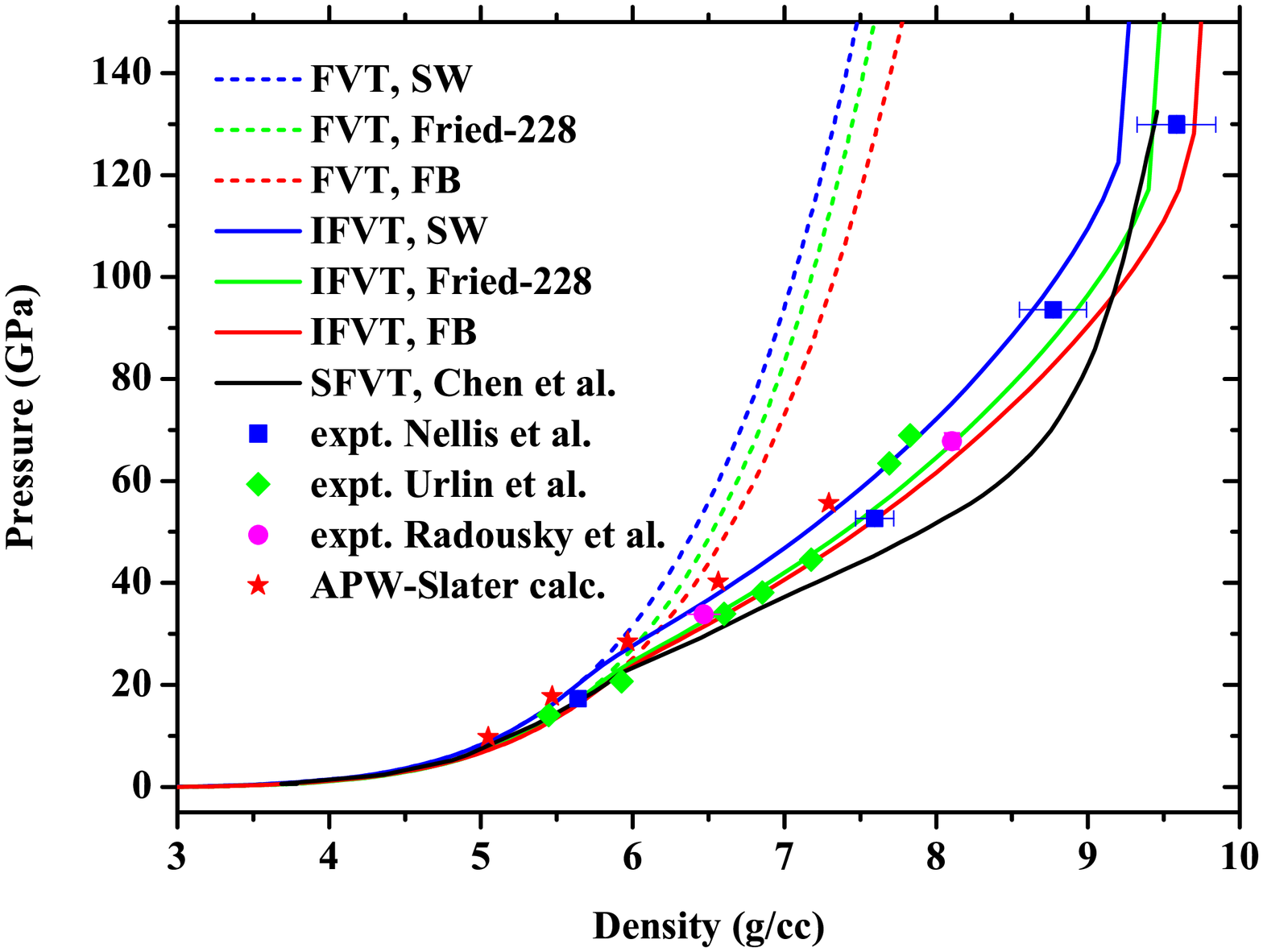}
\includegraphics[bb = 53 11 722 538,  width=8cm]{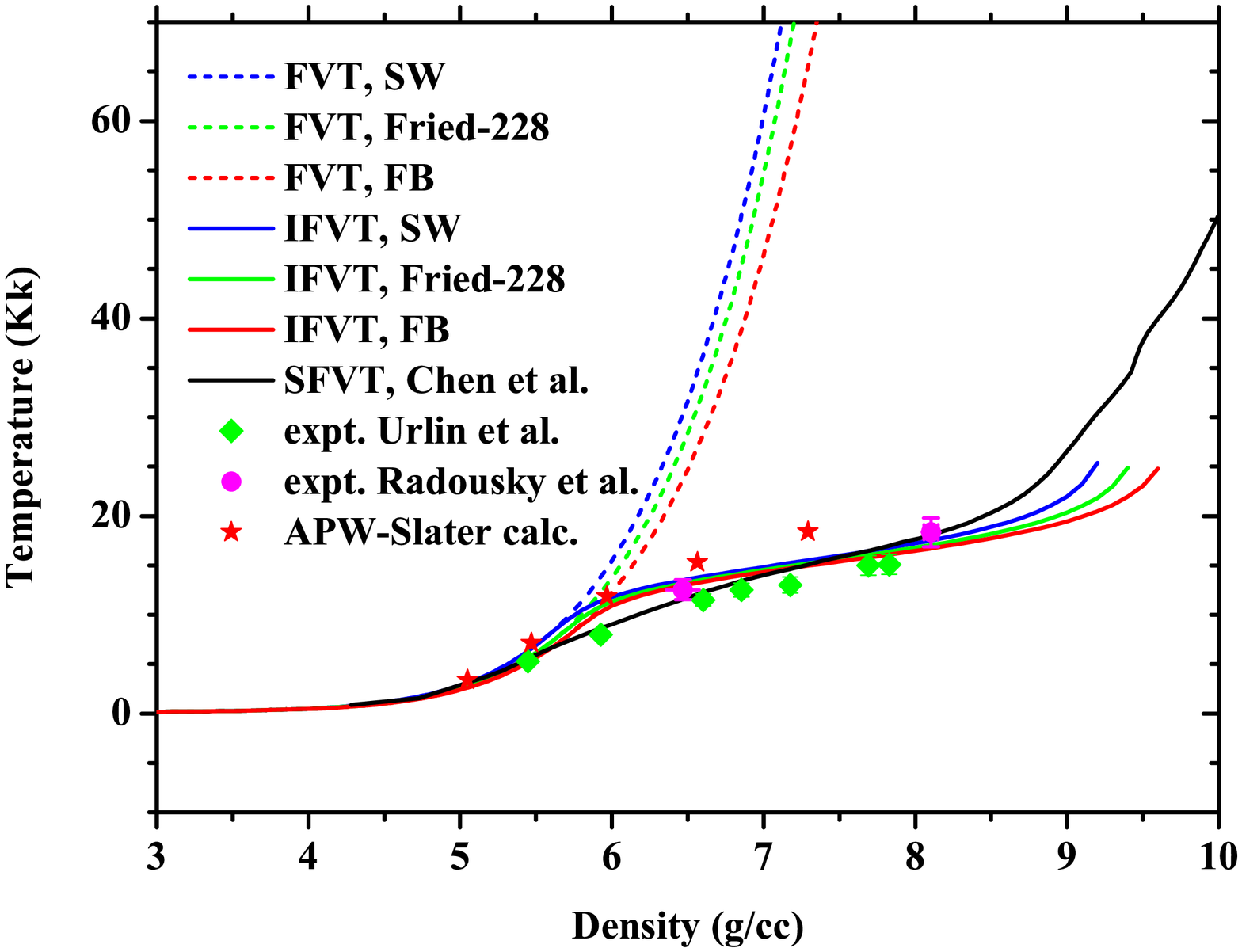}
\end{center}
\caption{\label{fig:NellisHugs} (Color online) The Hugoniot data of fluid xenon with the initial shock condition (165 K, 2.965 g/cc). The results calculated by the FVT model (short dash lines) and the IFVT model (solid lines)    using the pair potentials of "SW" (blue), "Fried-228" (green) and "FB" (red) respectively are compared. The SFVT chemical model calculation \cite{Chen2009} and the APW-Slater calculation \cite{1658} are also shown, together with the experimental shock data \cite{Nellis, Urlin, Radousky}. }
\end{figure*}

\begin{figure*}[btp]
\begin{center}
\includegraphics[bb = 53 11 722 538,  width=8cm]{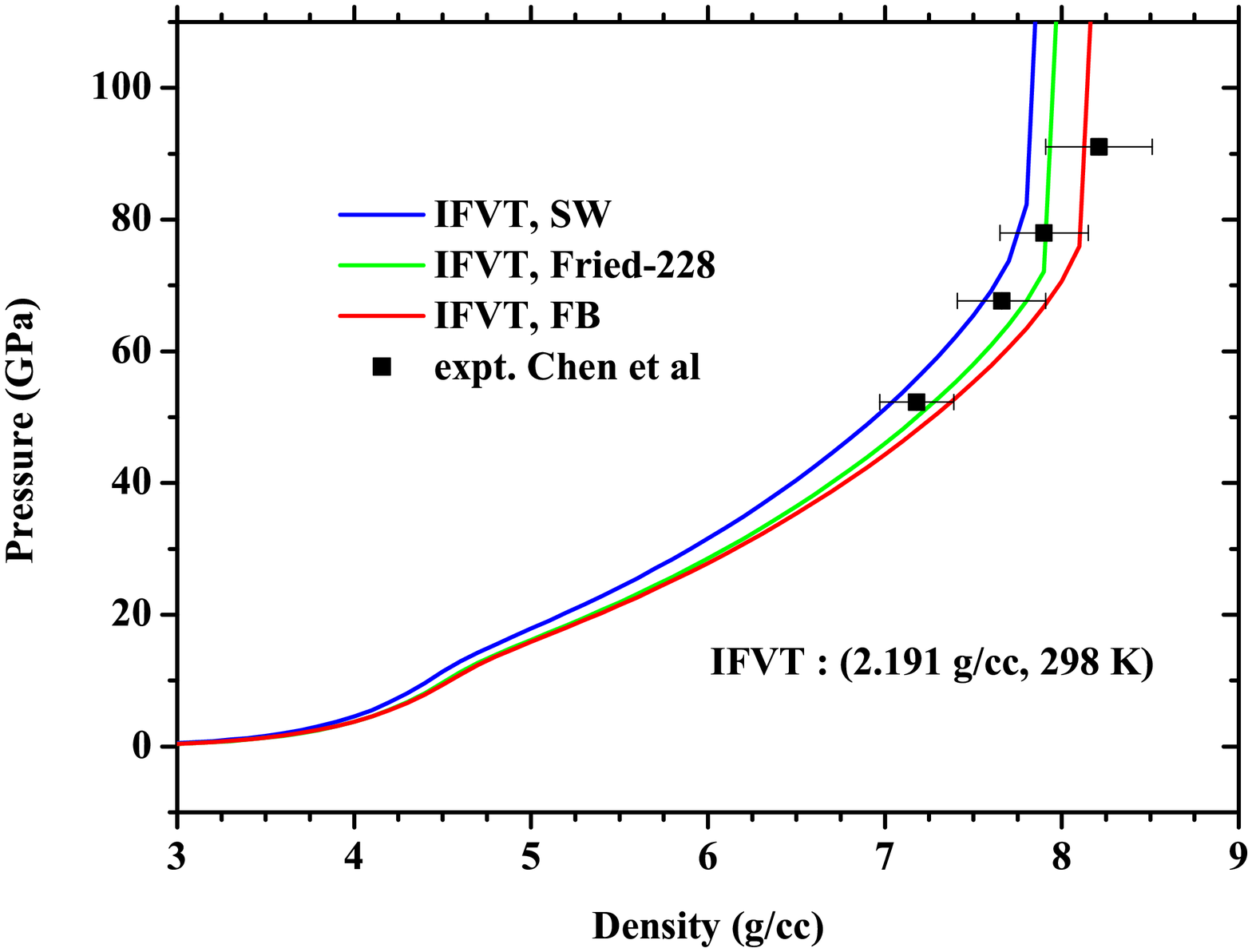}
\includegraphics[bb = 53 11 722 538,  width=8cm]{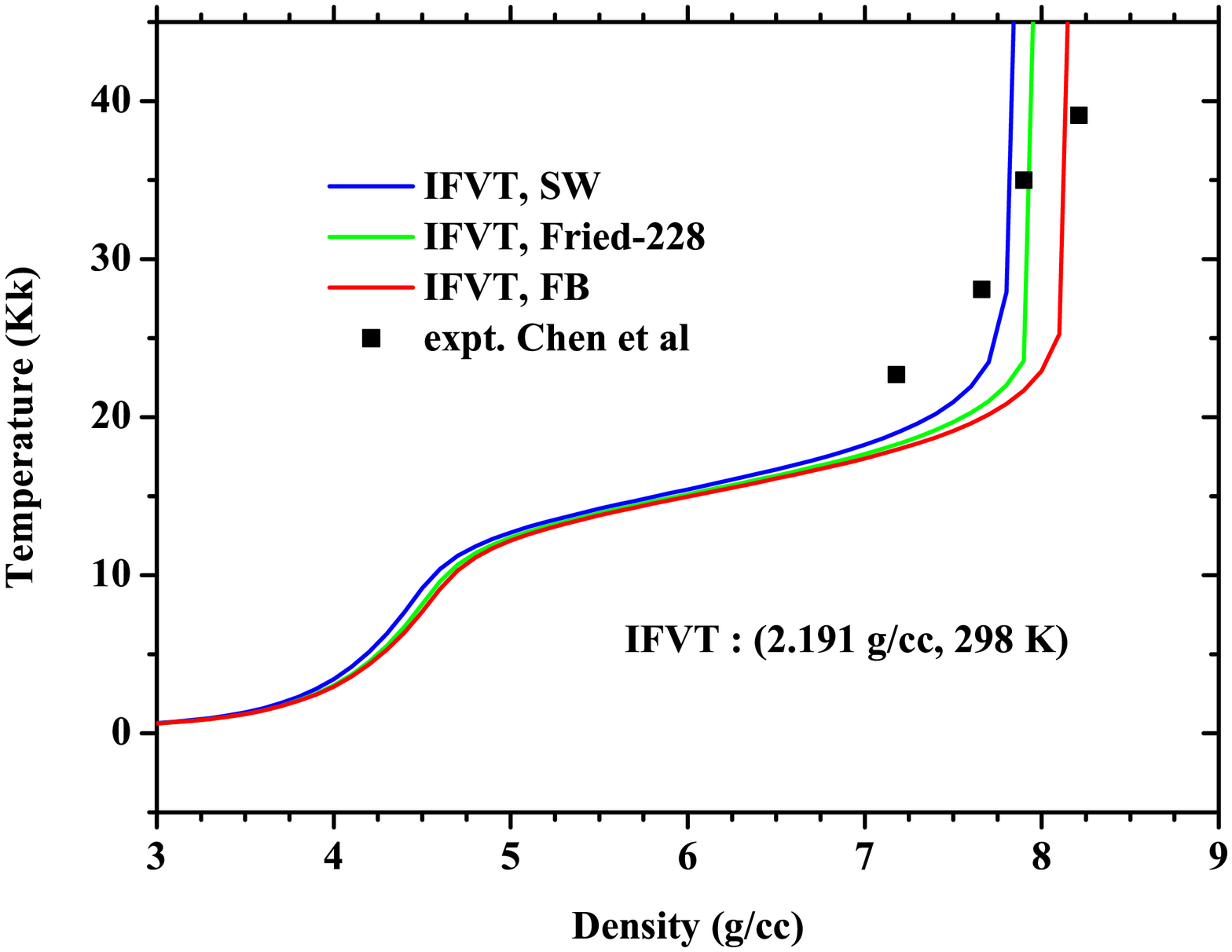}
\end{center}
\caption{\label{fig:ChenHugs} (Color online) The Hugoniot data of fluid xenon with the initial shock condition (298 K, 2.191 g/cc). The results calculated by the IFVT model (solid lines) using the pair potentials of "SW" (blue), "Fried-228" (green) and "FB" (red) respectively are compared with the experimental shock data \cite{Chen2012}. }
\end{figure*}

\section{Results}

This section demonstrates the results calculated by the FVT model and the IFVT model.

\subsection{FVT}

In the FVT model, the system is characterized by the inter-atomic pair potential, which should thus be determined properly. The xenon pair potentials of the exp-6 form available in the literature are plotted in Fig.\ref{fig:potential}. The "SW" potential, which is determined by the liquid argon shock wave data using the corresponding states theory \cite{Ross1980}, has been used in Ref.\cite{1658}. The "Fried-228" potential has been used in Refs.\cite{Schwarz2005, Chen2009}. The "Beam" potential, which is derived from the high-energy atomic-beam scattering data \cite{1658}, has not been used, because it is not in the form of an analytical function. We have fitted the "Beam" potential by the function of the exp-6 form, which is named as "FB". The parameters of the "FB" potential, the "SW" potential and the "Fried-228" potential are listed in Table \ref{table:par}.

The 300 K pressure-density relation of xenon is calculated by the FVT model using the pair potentials of "SW", "Fried-228" and "FB" respectively. As shown in Fig.\ref{fig:298K}, the FVT calculations are in reasonable agreement with the DFT calculation \cite{Root2010} and the experimental data \cite{Zisman,Jephcoat,Goettel}. At high densities, the results of the "SW" potential and the "Fried-228"  potential are higher than the experimental data  \cite{Goettel}, whereas the result of the "FB" potential, which is fitted by the "Beam" data, is in better agreement with the experimental data \cite{Goettel}.

The principal Hugoniot data of xenon is also calculated by the FVT model using the pair potentials of "SW", "Fried-228" and "FB" respectively. As shown in Fig.\ref{fig:NellisHugs}, the pressure and the temperature begin to deviate from the experimental shock data very quickly when the density increases up to 5.6 g/cc.

\subsection{IFVT}

In the IFVT model, electronic excitations and ionizations are taken into account, which can absorb part of the energy imparted to the system by shock waves and thus keep the temperature and the thermal pressure down.
Therefore, when electronic excitations and ionizations come into play, the Hugoniot curve of IFVT will bend over in comparison with that of FVT. As shown in Fig.\ref{fig:NellisHugs}, the Hugoniot curve of IFVT begins to bend over
from the point of about (5.6 g/cc, 6000 K, 20 GPa), and achieves good agreement with the available shock data \cite{Nellis,Urlin,Radousky}.

There are also other theoretical results shown in
Fig.\ref{fig:NellisHugs}. In the APW-Slater calculations
\cite{1658}, electronic transitions from 5p-like core states to
6s-like and 5d-like conduction bands are treated using semiconductor
statistics based on the augmented-plane-wave (APW)
electron-band-theory. It is shown that the APW-Slater Hugoniot data
is slightly higher than the shock data \cite{Nellis,Urlin,Radousky}.
In the SFVT model \cite{Chen2009}, electronic excitations and
ionizations are taken into account in the chemical picture. It is
shown that in the range $8\sim9$ g/cc the SFVT temperature is in
good agreement with the shock data, and the SFVT pressure is lower
than the shock data.

In Fig.\ref{fig:ChenHugs} the IFVT model is further compared with the experimental shock data of Zheng \emph{et al}. \cite{Chen2012}. It is shown that the Hugoniot pressure of IFVT is in good agreement with the experimental data, and the Hugoniot temperature of IFVT is lower than the experimental data.

\section{Summary and Outlook}

We have carried out an EOS study of fluid Xenon based on the fluid perturbation theory. It is demonstrated that the FVT model can be greatly improved by taking electronic excitations and ionizations into account in the framework of average ion spheres. Using the improved FVT model (IFVT), the EOS of fluid Xenon is calculated and achieves good agreement with the available shock data.

Note that the IFVT model uses the energy levels of isolated ions, which is a good approximation when electronic excitations and ionizations are weak enough.
An accurate calculation of the energy levels consistent with electronic excitations and ionizations will make the model more general.

\section{Acknowledgement}
This work was supported by the National Science Foundation of China
under Grants No.11604018, No.10804011 and No.11204015, by the
Foundation of LCP under Grants No.SNSYS16-027, and by the Foundation
under Grants No.JCKY2016212A501.

\end{document}